\documentclass[traditabstract]{aa} % for the abstract without structuration 
\usepackage{graphicx}
\usepackage{txfonts}

\begin{document}
 
% Title Page
\title{The masses and radii of HD186753B and TYC7096-222-1B: the discovery of two M-dwarfs that eclipse A-type stars}

\author{S. J. Bentley\inst{1}, B. Smalley\inst{1}, P. F. L. Maxted\inst{1}, C. Hellier\inst{1}, D. M. Wilson\inst{1,}\inst{2}, D. R. Anderson\inst{1}, A. Collier Cameron\inst{3}, M. Gillon\inst{5,}\inst{6}, L. Hebb\inst{3,}\inst{4}, D. L. Pollacco\inst{7}, D. Queloz\inst{5}, A. H. M. J. Triaud\inst{5}, R. G. West\inst{8}}

\authorrunning{S. J. Bentley et al.}
\titlerunning{The Masses and Radii of HD186753B and TYC7096-222-1B}

\institute{Astrophysics Group, Keele University, Keele, Staffordshire, ST5 5BG, U.K. \and
Centre for Astrophysics \& Planetary Science, School of Physical Sciences, University of Kent, Canterbury, Kent, CT2 7NH, UK \and
School of Physics and Astronomy, University of St. Andrews, North Haugh, Fife, KY16 9SS, U.K. \and
Vanderbilt University, Department of Physics and Astronomy, Nashville, TN 37235 \and
Observatoire de Gen\`{e}ve, Universit\'{e} de Gen\`{e}ve, 51 Chemin des Maillettes, 1290 Sauverny, Switzerland \and
Institut d'Astrophysique et de G\'{e}ophysique, Universit\'{e} de Li\`{e}ge, All\'{e}e du 6 Ao\^ut, 17, Bat. B5C, Li\`{e}ge 1, Belgium \and
Astrophysics Research Centre, School of Mathematics \& Physics, Queen’s University, University Road, Belfast, BT7 1NN, U.K. \and
Department of Physics and Astronomy, University of Leicester, Leicester, LE1 7RH, U.K.}

\date{}

\abstract {We present observations of two new single-lined eclipsing binaries, both consisting of an Am star and an M-dwarf, discovered by the Wide Angle Search for Planets transit photometry survey.  Using WASP photometry and spectroscopic measurements we find that HD186753B has an orbital period of $P=1.9194$ days, a mass of $M=0.24\pm0.02 M_{\odot}$ and radius of $R=0.31^{+0.06}_{-0.06} R_{\odot}$; and that TCY7096-222-1B has an orbital period of $P=8.9582$ days, a mass of between 0.29 and 0.54 $M_{\odot}$ depending on eccentricity and radius of $R=0.263^{+0.02}_{-0.07} R_{\odot}$.  We find that the Am stars have relatively low rotational velocities that closely match the orbital velocities of the M-dwarfs, suggesting that they have been ``spun-down'' by the M-dwarfs.}

\maketitle

\section{Introduction}
The radius of a star is one of its most fundamental properties, yet for sub-solar masses models have not been able to provide accurate radius predictions.  Citing the discrepancies between model and empirical radius measurements, Chabrier et al. (\cite{chabrier}) found that large surface spot coverage decreases the photospheric temperature.  The star compensates by increasing its radius to conserve radiative pressure.  This was confirmed by an empirical activity-radius study by Lopez-Morales (\cite{lopez-morales}).  In addition, Berger et al. (\cite{berger}) found a correlation between an increase in metalicity and a larger-than-expected radius.

Because of their low intrinsic brightness, low-mass stars (LMS) are particularly difficult to study. LMS in eclipsing binary systems (EBLM), however, provide a direct way to obtain radius measurements and are therefore a valuable tool for testing models of stellar structure in the low-mass region.  A by-product of wide-angle transit photometry planet-searching projects is the discovery of EBLMs (e.g. Fernandez et al. \cite{newtrespaper}).
  
The metallic-line Am stars are a class of peculiar A-type stars that are slow rotating, thought to have had their rotational velocity reduced by a near stellar companion.  Spectroscopic orbits of many Am stars have been reported (e.g. Carquillat \& Prieur \cite{ampaper}, Renson \& Manfroid \cite{renson}) and LMS are thought to be responsible for reducing the rotational velocity of Am stars (Carquillat \& Prieur \cite{ampaper}). Here we report the discovery of two single lined A-M binaries, HD186753 and TYC7096-222-1, the first EBLMs discovered from the Wide Angle Search for Planets (WASP) planet-hunting project.  There are four likely eclipsing A-M systems that have previously been announced; three of these systems were announced in Dreizler et al. (2002) and one in Pont et al. (2005).

\section{Observations}
\begin{figure}
 \centering
 \resizebox{\columnwidth}{!}{\includegraphics[angle=270,bb=50 50 554 770]{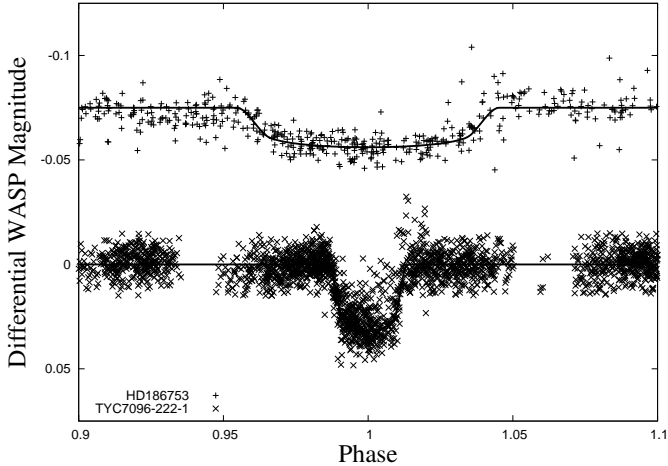}}
 % M_vs_R.eps: 1048592x1048592 pixel, 300dpi, 8878.08x8878.08 cm, bb=50 50 554 770
 \caption{\label{wasplc} The phase-folded WASP lightcurve of the eclipse of HD186753A (offset by 0.075 magnitudes) and TYC7096-222-1A overplotted with the best-fit models. The mean photometric uncertainty is 0.010 for HD186753 and 0.018 for TYC7096-222-1.}
\end{figure}

HD186753 (V=8.82) and TYC7096-222-1 (V=10.34) were identified to be eclipsing systems by analysis of time series photometry observed between May 2006 and May 2008 (HD186753) and May 2004 and March 2008 (TYC7096-222-1) by the WASP-South observatory, totalling 11,771 and 11,879 photometric measurements, respectively.  The WASP-South passband has an effective wavelength of 581 nm with a passband width of 225 nm, which is effectively a combination of the Johnson-Cousins V and I-bands.   Data were reduced with the WASP pipeline, de-trending and analysis tools described in Pollacco et al. (\cite{don}) and Collier Cameron et al. (\cite{mcmc}).  HD186753 showed a recurrent eclipse signature with a depth of 0.015 magnitudes and a period of 1.9194 days, whilst TYC7096-222-1 showed an eclipse depth of 0.024 magnitudes and a period of 8.9582 days, as shown in Fig. \ref{wasplc}.

\begin{figure}
 \centering
 \resizebox{\columnwidth}{!}{\includegraphics[angle=270,bb=50 50 554 770]{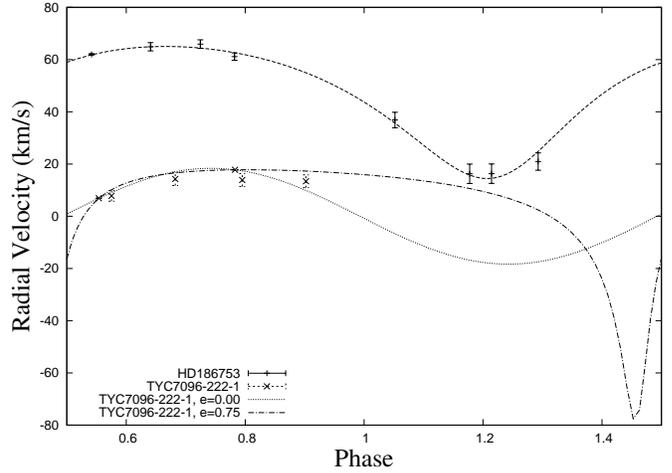}}
 % M_vs_R.eps: 1048592x1048592 pixel, 300dpi, 8878.08x8878.08 cm, bb=50 50 554 770
 \caption{\label{rv} The phase-folded radial velocity data of HD186753 (offset by 50 $\rm{km\ s^{-1}}$) and TYC7096-222-1 overplotted with the best-fit models.}
\end{figure}

In October 2008 we obtained radial velocity measurements of HD186753 and TYC7096-222-1 using the grating spectrometer on the 1.9-m telescope at the South African Astronomical Observatory (SAAO) at Sutherland, South Africa.  HD186753A has been identified by the SAO catalogue to be an A2m spectral type, whilst we used the InfraRed Flux Method (Blackwell \& Shallis \cite{irfm}) to estimate the spectral type of TYC7096-222-1A as an F0.  Because of this we used the spectral range $4666-4370$ \AA{} that contains the Mg~{\sc ii} 4481 \AA{} doublet, a spectral feature commonly used to obtain radial velocity measurements of early type stars.  Two consecutive spectra were coadded and we used the \textsc{Molly} package to determine, by cross-correlation with a synthetic template, the radial velocity of the system.  We obtained a single spectrum of HD186753 from the CORALIE spectrograph on the 1.2-m Leonhard Euler telescope and a single spectrum from the HARPS spectrograph on the 3.6-m telescope at La Silla, Chile, on the nights of 31st August 2008 and 7th October 2008, respectively.  We also obtained two spectra of TYC7096-222-1 from CORALIE on the nights of 31st August and September 11th 2008.  We used the radial velocity standard star HD8779 to calibrate the SAAO radial velocity data.  The radial velocity data are shown in Fig. \ref{rv} and Table \ref{rvdata}.

\begin{table}
% use packages: array
\begin{tabular}{lll}
HJD & RV & $\sigma_{RV}$ \\
-2,450,000 & ($\rm{km\ s^{-1}}$) & ($\rm{km\ s^{-1}}$) \\ \hline\hline
HD186753 & & \\ \hline
$4617.83677^{\rm{H}}$ & 5.20 & 0.28 \\
$4657.17198^{\rm{C}}$ & -19.96 & 2.95 \\
4762.28479 & -40.05 & 3.33 \\
4763.26196 & 0.15 & 1.41 \\
4764.34678 & -44.69 & 3.72 \\
4765.29219 & 4.94 & 1.62 \\
4766.34139 & -44.74 & 3.72 \\
4767.36517 & 3.92 & 1.57 \\ \hline
TYC7096-222-1 & & \\ \hline
$4710.87886^{\rm{C}}$ & 21.86 & 0.08 \\
$4721.88521^{\rm{C}}$ & 10.98 & 0.10 \\
4763.56489 & 17.49 & 2.46 \\
4764.51536 & 17.95 & 2.49 \\
4765.52730 & 18.36 & 2.52 \\
4766.49429 & 11.88 & 2.05 \\
\end{tabular}
\caption{\label{rvdata} Radial velocity measurements of HD186753 and TYC7096-222-1.  Superscript ``H'' denotes spectrum obtained by HARPS, ``C'' denotes CORALIE.}
\end{table}

\section{System Parameters}
For both targets we used the highest signal-to-noise (SNR) spectra taken by CORALIE ($\rm{\sim50:1}$) to estimate the stellar parameters of the primary. The low SNR has lead to relatively large errors on the final stellar parameters. Analysis of this spectrum was performed by the \textsc{uclsyn} spectral synthesis package (Smith \cite{uclsynsmith}; Smalley \cite{uclsynbarry}) and \textsc{atlas9} (Castelli et al. \cite{castelli}) models without convective overshooting.

The $H_{\alpha}$ and $H_{\beta}$ lines were used to determine the effective temperature, $\rm{T_{eff}}$. The equivalent widths of several clean and unblended lines were measured. Atomic line data was mainly taken from the Kurucz \& Bell (1995) compilation, but with updated van der Waals broadening coefficients for lines in Barklem et al. (2000) and $\log{gf}$ values from Gonzalez \& Laws (2000), Gonzalez et al. (2001) or Santos et al. (2004). The ionization balance between Fe~{\sc i} and Fe~{\sc ii} was used to determine surface gravity, $\log{g}$. A value for microturbulence, $\xi_{\rm{t}}$, was determined from Fe~{\sc i} using Magain's (1984) method. Elemental abundances were determined from their measured equivalents. The quoted error estimates include that given by the uncertainties in $\rm{T_{eff}}$, $\log{g}$ and $\xi_{\rm{t}}$, as well as the scatter due to measurement and atomic data uncertainties. The projected stellar rotation velocity ($v\sin{i}$) was determined by fitting the profiles of several unblended Fe~{\sc i} lines.

For HD186753A this fit yielded $\rm{T_{eff}}=7800\pm200~\rm{K}$, $\log{g}=4.0\pm0.3$, $v\sin{i}=65.0\pm5.0\rm{km\  s^{-1}}$ and $\xi_{\rm{t}}=4.0\pm0.5\rm{km\ s^{-1}}$.  HD186753A is listed as having a spectral type A2mA5-F0 in Houck \& Smith-Moore (1988). Our analysis is consistent with this classification.   Using the derived $\rm{T}_{\rm{eff}}$ from the spectrum and the stellar density, $\rho_{1}$, from the lightcurve analysis we used the isochrones of Girardi et al. (\cite{girardi}) to estimate a stellar mass of $M_{1}=1.79\pm{0.05}M_{\odot}$ and WASP photometry to determine the radius as $R_{1}=2.53^{+0.27}_{-0.30}R_{\odot}$.  For TYC7096-222-1A we find $\rm{T_{eff}}=7600\pm300$ K, $\log{g}=4.0\pm0.3$, $v\sin{i}=35.0\pm5.0\rm{km\  s^{-1}}$ and $\xi_{\rm{t}}=4.0\pm0.5\rm{km\ s^{-1}}$.  Using the derived $\rm{T_{eff}}$ and $\rho_{1}$ values with the Girardi et al. (\cite{girardi}) isochrones we find a mass of $M_{1}=1.74\pm0.05M_{\odot}$ and radius of $R_{1}=1.66^{+0.10}_{-0.08}R_{\odot}$.  The mass and radius of each primary is consistent within $1\sigma$ of the empirical mass-radius measurements of A-type stars discussed in Torres et al. (\cite{torresetal}).  The results of the fit are shown in Table \ref{stellarparams}.

\begin{table}
% use packages: array
\begin{tabular}{lll}
Parameter & HD186753A & TYC7096-222-1A \\ \hline\hline
Stellar Mass, $M_{1}(M_{\odot})$ & $1.794\pm{0.046}$ & $1.735\pm0.054$ \\ 
Stellar Radius, $R_{1}(R_{\odot})$ & $2.527^{+0.270}_{-0.303}$ &  $1.662^{+0.103}_{-0.081}$\\
Stellar Density, $\rho_{1} (\rho_{\odot})$ & $0.367^{+0.119}_{0.078}$ & $0.375^{+0.052}_{-0.051}$ \\ 
$\rm{T_{eff}} (\rm{K})$ & $7800\pm200$ & $7600\pm300$ \\ 
$\log{g}$ & $4.0\pm0.3$ & $4.0\pm0.3$ \\ 
$v\sin{i} (\rm{km\ s^{-1}})$ & $65.0\pm5.0$ & $35.0\pm5.0$ \\ 
$\xi_{\rm{t}} (\rm{km\ s^{-1}})$ & $4.0\pm0.5$ & $4.0\pm0.5$ \\ 
$\rm{[Fe/H]}$ & $+0.12\pm0.12$ & $+0.08\pm0.13$ \\
$\rm{[Ca/H]}$ & $-0.41\pm0.14$ & $-0.63\pm0.09$ \\
$\rm{[Sc/H]}$ & $-0.65\pm0.12$ & $-1.47\pm0.05$ \\
$\rm{[Y/H]}$ & $+0.55\pm0.13$ & $+0.54\pm0.16$ \\
$\rm{[Ba/H]}$ & $+0.87\pm0.16$ & $+1.20\pm0.22$ \\
\end{tabular}
\caption{\label{stellarparams} Parameters of HD186753A and TYC7096-222-1A.}
\end{table}

We used the \textit{Systemic Console} (Meschiari et al. \cite{systemic}) to fit our radial velocity data.  The period was fixed to the period determined from WASP photometry, with phase=1 set at the centre of the eclipse.  In our fitting procedure we assumed a Keplerian orbit of the M-dwarfs and fitted the stellar reflex velocity, $\rm{K}_{1}$, the centre-of-mass velocity, $\gamma$ and for HD186753 we also fitted the eccentricity, $e$, and the longitude of periastron, $\omega$.  The phase coverage of radial velocity data for TYC7096-222-1 is not sufficient for an accurate value of $e$.  We therefore fixed $e$ to 0.75, which is the highest eccentricity for an Am binary with a period between 4-12 days (Carquillat \& Prieur \cite{ampaper}) and left the remaining parameters free to fit to see what affect a high value of $e$ has on the secondary mass, $M_{2}$.  The lightcurves were anylysed with the MCMC code described in Collier Cameron et al. (\cite{mcmc}).  The M-dwarfs were analysed as though they are dark objects as they both contribute to only 0.06\% of light in the WASP passband.

We estimated the mass of the secondaries by using the inclination estimated from the MCMC analysis of the photometry and $K_{1}$ from the \textit{Systemic Console} analysis of the radial velocity data. HD186753B has a mass of $M_{2}=0.24\pm0.02 M_{\odot}$ and radius of $R_{2}=0.31^{+0.06}_{-0.06} R_{\odot}$.  TYC7096-222-1B has a mass of  $M_{2}=0.29\pm0.02 M_{\odot}$ and radius of $R_{2}=0.31^{+0.06}_{-0.06} R_{\odot}$, assuming zero eccentricity. Setting $e=0.75$ gives TYC7096-222-1B a mass of $M_{2}=0.54\pm0.06 M_{\odot}$.  The results of the fit are shown in Table \ref{paramtable}.

\section{Discussion}\label{discussion}
The rotational velocity of A8 stars is expected to be $\sim~200\rm{km\ s^{-1}}$ (Gray \cite{gray}).  The relatively low rotational velocities of HD186753A and TYC7096-222-1A, $v\sin{i}=65.0\pm5.0$ and $35.0\pm5.0 \rm{km\ s^{-1}}$, respectively, suggests that they have been ``spun-down'' by their M-dwarf companions.  The stellar rotational angular velocity of HD186753A is $(3.67\pm0.65)\times10^{-4} \rm{rad\ s^{-1}}$ which is greater than the M-dwarf orbital angular velocity of $(3.79\pm0.01)\times10^{-5} \rm{rad\ s^{-1}}$. The rotational angular velocity of TYC7096-222-1A is $(3.16\pm0.47)\times10^{-5} \rm{rad\ s^{-1}}$ which is also greater than the M-dwarf orbital angular velocity of $(8.12\pm0.02)\times10^{-6} \rm{rad\ s^{-1}}$.  The synchronisation timescales are 2.8 Myr for HD186753 and 0.92 Gyr for TYC7096-222-1 (Zahn 1977), suggesting that each system is younger than the synchronisation time.  Values of $\log{g}$ can be used as an age estimate, but our value is not reliable enough.  Higher SNR spectra is required to determine the age of the systems.

The eccentricity of HD186753B, $e=0.269\pm0.087$, is quite high for a short period binary and could also indicate a young stellar age as the circulization time is 95 Myr (Zahn 1977).  The circulization time for  TYC7096-222-1B is 295 Gyr.  If TYC7096-222-1B has a mass of $M_{2}=0.54\pm0.06 M_{\odot}$ (using the $e=0.75$ model) then we would expect a secondary eclipse of $0.003\pm1\times{}10^{-4}$ mag deep.  There appears to be no sign of this in the WASP photometry at any phase.  TYC7096-222-1B therefore has a mass between $0.29-0.54 M_{\odot}$ depending on the eccentricity. The certainty of $e$ for both objects could be improved with greater radial velocity phase coverage.  The value of $e$ for HD186753B could be explained by a tertiary component either in the system or by a recent near-miss, although we find no evidence in our data for either scenario.

Both primaries show supersolar Fe abundances, an underabundance of Ca and Sc and overabundances of Y and Ba. These abundances are typical of Am stars (e.g. Wolff \cite{wolff}; Hundt \cite{hundt}).  Carquillat \& Prieur (\cite{ampaper}) found that the mean companion mass to an Am spectral type is $0.8\pm{0.5}M_{\odot}$ with a mean orbital period of 1.995 days. By these stellar properties HD186753 and TYC7096-222-1 are fairly typical Am systems.

\begin{figure}
 \centering
 \resizebox{\columnwidth}{!}{\includegraphics[angle=270,bb=50 50 554 770]{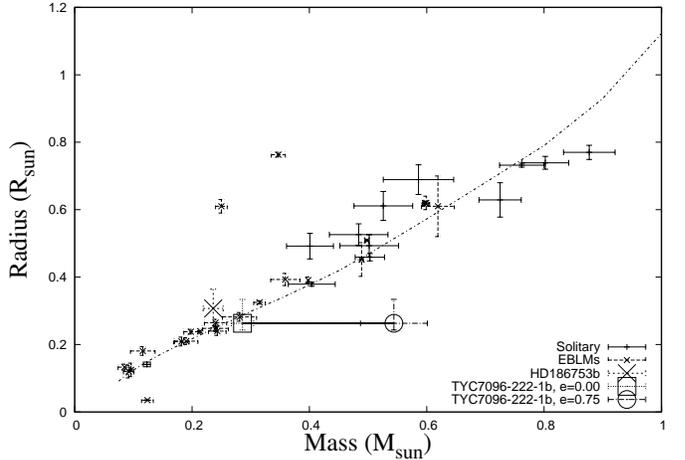}}
 % M_vs_R.eps: 1048592x1048592 pixel, 300dpi, 8878.08x8878.08 cm, bb=50 50 554 770
 \caption{Mass-radius relationship for solitary stars, EBLMs, HD186753B and TCY7096-222-1 superimposed on 5 Gyr, [M/H]=0, $L_{\rm{mix}}=1H_{p}$ Baraffe et al. (\cite{baraffe}) isochrones. The thick black line indicates the range of masses for TYC7096-222-1B depending on the eccentricity. M-dwarfs with uncertainties in mass $>0.03 M_{\odot}$ have been omitted from the figure for illustrative clarity.}
 \label{m-r}
\end{figure}

The radii of stars with masses below $0.3-0.35\rm{M_{\odot}}$ agree well with the Baraffe et al. (\cite{baraffe}) isochrone models (Lopez-Morales \cite{lopez-morales}).  The mass-radius relation TYC7096-222-1B agree within $1\sigma$ of the Baraffe et al. (\cite{baraffe}) isochrones, as shown in Fig. \ref{m-r}. Whilst the mass-radius relation of HD186753B, however, is only just within $1\sigma$ of the Baraffe et al. (\cite{baraffe}) isochrones.  M-dwarfs in binaries are found to be more active than solitary M-dwarfs (Lopez-Morales \cite{lopez-morales}). Increased activity causes a decrease in photospheric temperature, which then causes an increase in radius to conserve radiative pressure (Ribas \cite{Ribas}).  X-ray activity is an indicator of stellar activity, although no X-ray data has been published on either object.

Being in a tight orbit, the photosphere of HD186753B is strongly irradiated by HD186753A.  The radiation intensity received from HD186753A at the photosphere of HD186753B, and also therefore the radiative pressure, is $1.32\pm0.02$ times the radiation generated; the value for TYC7096-222-1 is lower at $0.72\pm0.04$.  Higher precision radial velocity and photometry than the values given here are required to ascertain as to whether this extra source of heating, or increased levels of activity are affecting the mass-radius relation of the M-dwarfs.

%\begin{figure}
% \centering
% \resizebox{\columnwidth}{!}{\includegraphics[angle=270,bb=50 50 554 770]{Residuals.ps}}
% % M_vs_R.eps: 1048592x1048592 pixel, 300dpi, 8878.08x8878.08 cm, bb=50 50 554 770
% \caption{Residuals from the mass-radius relationship shown in Fig. \ref{m-r}.}
% \label{m-r_resid}
%\end{figure}

\begin{table*}
% use packages: array
\begin{tabular}{llll}
 Parameter & HD186753B & TYC7096-222-1B ($e=0.00$) & TYC7096-222-1B ($e=0.75$) \\   \hline\hline
Eclipse epoch (HJD) & $3940.40144^{+0.00091}_{-0.00081}$ & $4373.01637^{+0.00066}_{-0.00087}$ & - \\
Orbital period (days) & $1.9193851^{+0.0000412}_{-0.0000393}$ & $8.9582591^{+0.0000346}_{-0.0000314}$ &  - \\
Eclipse duration (days) & $0.1662^{+0.0038}_{-0.0031}$ & $0.2430^{+0.0039}_{-0.0042}$  &  - \\
Secondary/primary area ratio, $(R_{2}/R_{1})^2$ & $0.0148^{+0.0005}_{-0.0003}$ & $0.0251^{+0.0006}_{-0.0005}$ &  - \\
Impact parameter, $b (R_{*})$ & $0.264^{+0.191}_{-0.154}$ & $0.250^{+0.157}_{-0.142}$  &  - \\
Stellar reflex vel., $\rm{K_{1}}$ ($\rm{km\ s^{-1}}$) & $-27.449\pm1.751$ & $-20.341\pm2.974$  & $-58.432\pm6.122$ \\
Centre-of-mass vel., $\gamma$ ($\rm{km\ s^{-1}}$) & $-14.641\pm2.980$ & $4.074\pm1.533$  & $8.776\pm0.748$ \\
Orbital separation, $a$ (AU) & $0.0370^{+0.0012}_{-0.0013}$ & $0.0990^{+0.0031}_{-0.0034}$  &  - \\
Orbital inclination, $i$ (deg) & $87.09^{+1.69}_{-1.79}$ & $89.00^{+0.57}_{-0.73}$  &  - \\
Orbital eccentricity, $e$ & $0.269\pm0.087$ & (fixed=0.00)  & (fixed=0.75) \\
Longitude of periastron, $\omega$ (deg) & $166.7\pm5.9$ & (not fitted)  & $167.2\pm61$ \\
Stellar Mass, $M_{2}(M_{\odot})$ & $0.236\pm0.017$ & $0.286\pm0.019$  & $0.544\pm0.057$ \\
Stellar Radius, $R_{2}(R_{\odot})$ & $0.307^{+0.057}_{-0.057}$ & $0.263^{+0.020}_{-0.071}$  &  - \\
Luminosity ratio, $L_{1}/L_{2}$ & $979.7$ & $520.4$  & $361.9$ \\
\end{tabular}
\caption{\label{paramtable} Parameters of HD186753B and TYC7096-222-1B and their orbits.  The parameters of TYC7096-222-1B determined from the photometry are the same for both eccentricities.}
\end{table*}

\begin{acknowledgements}
SJB acknowledges the support of an STFC postgraduate studentship.  We are very greatful to the A\&A referee Jonathan Devor for his patience, assistance and helpful suggestions. The WASP consortium comprises the Universities of Keele, Leicester, St. Andrews, the Queen’s University Belfast, the Open University and the Isaac Newton Group. WASP-South is hosted by the South African Astronomical Observatory and we are grateful for their support and assistance. Funding forWASP comes from consortium universities and from the UK’s Science and Technology Facilities Council. We also thank Tom Marsh for the use of his \textsc{Molly} spectra reduction programme.
\end{acknowledgements}

\end{document}